\documentclass[12pt]{article}

\usepackage{epsfig}
\usepackage{graphicx}
\usepackage{amssymb,amsmath}
\usepackage{hyperref}
\hypersetup{
    colorlinks=true,
    linkcolor=blue,
    citecolor=red,
}

\renewcommand{\Re}{\text{Re}\,}
\begin{document}

\begin{center}
\Large{\bf Lorentzian Vacuum Transitions in Ho\v{r}ava-Lifshitz
Gravity} \vspace{0.5cm}

\large  H. Garc\'{\i}a-Compe\'an\footnote{e-mail address: {\tt
compean@fis.cinvestav.mx}}, D. Mata-Pacheco\footnote{e-mail
address: {\tt dmata@fis.cinvestav.mx}}

\vspace{0.3cm}

{\small \em Departamento de F\'{\i}sica, Centro de
Investigaci\'on y de Estudios Avanzados del IPN}\\
{\small\em P.O. Box 14-740, CP. 07000, Ciudad de M\'exico, M\'exico}\\

\vspace*{1.5cm}
\end{center}

\begin{abstract}
The vacuum transition probabilities for a
Friedmann-Lema\^itre-Robertson-Walker universe with positive
curvature in Ho\v{r}ava-Lifshitz gravity in the presence of a scalar
field potential in the Wentzel-Kramers-Brillouin approximation are
studied. We use a general procedure to compute such transition
probabilities using a Hamiltonian approach to the Wheeler-DeWitt
equation presented in a previous work. We consider two situations of
scalar fields, one in which the scalar field depends on all the
spacetime variables and other in which the scalar field depends only
on the time variable. In both cases analytic expressions for the
vacuum transition probabilities are obtained and the infrared and ultraviolet
limits are discussed for comparison with the result obtained by
using general relativity. For the case in which the scalar field
depends on all spacetime variables we obtain that in the infrared
limit it is possible to obtain a similar behavior as in general
relativity, however in the ultraviolet limit the behavior found is
completely opposite. Some few comments about possible phenomenological implications 
of our results are given. One of them is a plausible resolution of the initial singularity. On the other hand for the case in which the
scalar field depends only on the time variable, the behavior
coincides with that of general relativity in both limits, although
in the intermediate region the probability is slightly altered.
\vskip 1truecm

\end{abstract}

\bigskip

\newpage

\section{Introduction}
\label{S-Intro}

The quantum theory of the gravitational phenomena, or quantum
gravity,  is a theory in construction which is necessary in order to
shed light about the quantum effects of gravitational systems. Among
the problems that require the uses of quantum gravity is the study
of the microscopic origin of thermodynamic properties of black
holes and those describing some cosmological phenomena in the very
early universe. Another important problem is the study of the vacuum
decay and the transition between vacua at early stages of the
evolution of the universe. Euclidean methods have been proposed in
order to compute this transition probability  by using the path
integral approach \cite{ColemanFTone,ColemanFTtwo,ColemanDeLuccia}.
One of the salient features of this approach is the prediction of
transitions between open universes \cite{ColemanDeLuccia}. Later an
alternative procedure to compute these transitions using the
Hamiltonian approach was developed \cite{FMPone,FMPtwo}. This method
incorporates the Arnowitt, Deser and Misner (ADM) Hamiltonian
formalism of general relativity (GR)
\cite{Arnowitt:1962hi,Wheeler,DeWitt}. The vacuum is implemented
through a cosmological constant which is interpreted as the vacuum
energy and the transition are carried out through a bubble
nucleation \cite{deAlwis:2019dkc}. In this approach the transitions
between Minkowski and de Sitter spaces are allowed. Very recently
approach \cite{FMPone,FMPtwo} was further developed by Cespedes et
al. \cite{Cespedes:2020xpn}, where the vacuum is implemented by the
minima of a potential of a scalar field in the curved space. In this
reference it was computed the general vacuum decay transitions in
the Hamiltonian formalism in Wheeler's superspace and some examples
were implemented in the minisuperspace formalism for the
Friedmann-Lema\^itre-Robertson-Walker (FLRW) cosmology. In this kind
of models it was shown that the transitions between closed universes
are allowed contrary to the Euclidean approximation of Coleman and
De Luccia \cite{ColemanDeLuccia}. Later the formalism of
\cite{Cespedes:2020xpn} was extended and used to obtain the vacuum
decay transition probabilities for some examples of transitions
between anisotropic universes \cite{LVTAU}.


On the other hand it is well known that GR is not a renormalizable
theory. Thus its application to very small distances as those
associated to the early universe is expected to fail. Instead of
that an important proposal to describe quantum effects of gravity is
the Ho\v{r}ava-Lifshitz (HL) theory \cite{Horava:2009uw} (for some
recent reviews, see
\cite{Weinfurtner:2010hz,Sotiriou:2010wn,Wang:2017brl,Mukohyama:2010xz}
and references therein).  Ho\v{r}ava-Lifshitz theory is a theory
with an anisotropic scaling of spacetime and consequently it is not
Lorentz invariant at high energies (ultraviolet (UV)). However it is
a well behaved description at small distances due to the
incorporation of higher order derivative terms in the spatial
components of the curvature to the usual Einstein-Hilbert action
giving rise to a ghost-free theory. Thus this theory is more
appropriate to describe the quantum effects of the gravitational
field, as the vacuum decay processes in the early stages of the
universe evolution.

It is important to remark that HL theory is a theory whose low
energy limit which connects with GR is troublesome. The parameters
of the theory are the critical exponent $z$ and the foliation
parameter $\lambda$. This last parameter is associated with a
restricted foliation compatible with the Lifshitz scaling. In the
low energy limit $z \to 1$ the Lorentz invariance is recovered. In
the infrared (IR) limit the $z \to 1$ limit is accompanied of the
limit $\lambda \to 1$, where the full diffeomorphisms symmetry is
recovered and consequently the usual foliation of the ADM formalism
is regained. In addition the higher order derivative terms in the
action have to be properly neglected in order to get the correct
limit. As we mentioned before the GR limit is problematic since it
remains an additional degree of freedom (in some cases interpreted
as dark matter) which leads to a perturbative IR instability
\cite{Sotiriou:2010wn,Mukohyama:2010xz,Izumi:2011eh,Gumrukcuoglu:2011ef}.
The non-projectable version of the HL theory has the possibility to
remove this unphysical degree of freedom. Thus it represents an
advantage over the projective theory. However in the case in which
one is concern with the Wheeler-DeWitt (WDW) equation, both
approaches give the same result. In consequence we will work with
the projectable version. In order to deal with these infrared instabilities
several alternative proposals has been presented such as in \cite{Cognola:2016gjy,Zhu:2011xe}, the former one was tested against gravitational waves measurements \cite{Casalino:2018wnc}. 

Since HL theory represents an improvement over GR in the high energy
regime, it is natural that quantum gravity aspects of the theory are
of great interest. Indeed, canonical quantization of the theory has
been extensively studied. For example, some of the papers describing
solutions of Ho\v{r}ava-Lifshitz's gravity in quantum cosmology in
the minisuperspace are
\cite{Bertolami:2011ka,Christodoulakis:2011np,Pitelli:2012sj,Vakili:2013wc,Obregon:2012bt,Benedetti:2014dra,Cordero:2017egl}.

As we mentioned before HL gravity is a UV completion of GR thus it
is a more suitable arena where to study the vacuum transitions in
the presence of a scalar field potential. This proposal will be
carried out in the present article. In order to do that we use the
Hamiltonian formalism of the HL theory, in particular the WDW
equation will be discussed in this context following
\cite{Cespedes:2020xpn,LVTAU}. We will particularly focus on the
closed FLRW universe and study two types of scalar fields. First,
since the anisotropic scaling of spacetime variables is a key
ingredient of HL theory, we will consider a scalar field which is
allowed to depend on all spacetime coordinates. Lastly we will
consider also a scalar field which only depends on the time variable
as it is more usual on the cosmological models.

This work is organized as follows. In Section \ref{S-WDW} we give a
brief review of the general procedure presented in \cite{LVTAU} to
study vacuum transition probabilities between two minima of a scalar
field potential in the minisuperpace following the formalism of
\cite{Cespedes:2020xpn}. We will show that this formalism
implemented for GR in \cite{Cespedes:2020xpn,LVTAU} is sufficient to
study vacuum transitions in a more general theory as HL theory.
Section \ref{S-HL} is devoted to obtain the WDW equation in the
context of gravity coupled to matter. In Section \ref{S-Trans} we
study the vacuum transitions in HL gravity for the scalar field
depending on all spacetime variables. The IR and UV limits for the
transition probabilities are discussed and compared to the GR
result. Then, in Section \ref{S-TT} we study the transition
probabilities for the scalar field depending only on the time
variable and we also compare the result to the GR one. Finally, in
Section \ref{S-FinalR} we give our conclusions and final remarks.

\section{Vacuum transitions for a scalar field}
\label{S-WDW}

In this section we review the procedure to obtain a general
expression for the transition probability between two minima of the
potential of a scalar field by obtaining a semi-classical solution
to the WDW equation using a WKB ansatz described in Ref.
\cite{LVTAU}. We follow closely the notation and conventions given
in that reference.  It is a remarkable point to see that this
procedure is enough to implement theories more general than GR as
the HL gravity.

We start by using the well known ADM-formulation of GR
\cite{Arnowitt:1962hi,Wheeler,DeWitt} and consider the Hamiltonian
constraint expressed in the general form
    \begin{equation}\label{HamConstraint}
        \mathcal{H}=\frac{1}{2}G^{MN}(\Phi)\pi_{M}\pi_{N}+f[\Phi]\approx0 ,
    \end{equation}
where we take the coordinates in Wheeler's superspace to be $\Phi^M$
with $M,N=1,...,n$ (which has in general an infinite number of dimensions).
These variables are the components of the three-dimensional metric,
the matter field variables, etc. and are denoted collectively
as $\Phi$. Their corresponding canonical momenta are $\pi_{M}$ and
the inverse metric in such space is $G^{MN}$. Finally, $f[\Phi]$ is
a function that represents all other additional terms as the $^3R$
term and the potential terms of scalar field in the WDW equation.
The general WDW equation that we are going to consider is obtained
after carrying out the standard canonical quantization procedure of
the Hamiltonian constraint. Thus performing this procedure we get
    \begin{equation}\label{WDWEQ}
        \mathcal{H}\Psi(\Phi)=\left[-\frac{\hbar^2}{2}G^{MN}(\Phi)\frac{\delta}{\delta\Phi^M}\frac{\delta}{\delta\Phi^N}+f[\Phi]\right]\Psi[\Phi]=0 ,
    \end{equation}
where $\Psi[\Phi]$ represents the wave functional which depends on
all fields of the theory.

We are interested in obtaining a semi-classical result, therefore
following \cite{Cespedes:2020xpn,LVTAU} we consider an ansatz of the
following WKB form
$\Psi[\Phi]=\exp\left\{\frac{i}{\hbar}S[\Phi]\right\},$ where $S$
has an expansion in $\hbar$ in the usual form
    \begin{equation}\label{SemiClasExp}
        S[\Phi]=S_{0}[\Phi]+\hbar S_{1}[\Phi]+\mathcal{O}(\hbar^2) .
    \end{equation}
Inserting Eq. (\ref{SemiClasExp}) into Eq. (\ref{WDWEQ}) and
focusing only on the term at the lowest order in $\hbar$ we obtain
    \begin{equation}\label{WDWOrder0}
            \frac{1}{2}G^{MN}\frac{\delta S_{0}}{\delta\Phi^{M}}\frac{\delta S_{0}}{\delta\Phi^{N}}+f[\Phi]=
            0.
    \end{equation}
On a certain slice of the space of fields a set of integral curves can
be specified in the form
    \begin{equation}\label{DefCs}
        C(s)\frac{d\Phi^M}{ds}=G^{MN}\frac{\delta
        S_{0}}{\delta\Phi^N},
    \end{equation}
where $s$ is the parameter of these curves. The classical
action appearing in the previous equation has the form
      \begin{equation}\label{ClassActionGen}
        S_{0}[\Phi_{s}]=-2\int^s\frac{ds'}{C(s')}\int_{X}f[\Phi_{s'}] .
    \end{equation}
It is easy to see that Eqs. (\ref{WDWOrder0}) and (\ref{DefCs})
leads to
    \begin{equation}\label{RelDerC2}
        G_{MN}\frac{d\Phi^M}{ds}\frac{d\Phi^N}{ds}=-\frac{2f[\Phi_{s}]}{C^2(s)} ,
    \end{equation}
where $G_{MN}$ satisfies the standard relation $G_{PM}G^{MN}=
\delta^N_P$.

We note that we have a system of equations for the $n+1$ variables:
$\left(\frac{d\Phi^M}{ds},C^2(s)\right)$ defined by (\ref{DefCs})
and (\ref{RelDerC2}). Thus we can obtain a solution for such a
system and then substitute the results back into Eq.
(\ref{ClassActionGen}) to obtain the classical action. Thus, in
principle, we have enough information to compute the classical
action, and consequently the wave functional to first order in
$\hbar$ regardless of the number of fields in superspace.

Under the ansatz that all the fields $\Phi^M$ on the superspace
depend only on the time variable, we can obtain a general solution to the system in terms of the volume Vol$(X)$ of the spatial slice $X$ of
the form
    \begin{equation}\label{CsGeneral}
        C^2(s)=-\frac{2{\rm Vol}^2(X)}{f[\Phi]}G^{MN}\frac{\partial f}{\partial\Phi^M}\frac{\partial f}{\partial\Phi^N} ,
    \end{equation}
    \begin{equation}\label{DerivSolGen}
            \frac{d\Phi^M}{ds}=\frac{f[\Phi]}{{\rm Vol}(X)}\frac{G^{MN}\frac{\partial f}{\partial\Phi^N}}{G^{LO}\frac{\partial f}{\partial\Phi^L}\frac{\partial f}{\partial\Phi^{O}}}
            .
    \end{equation}

In this article we will consider gravity coupled to a scalar field
provided of a potential which have at least a false and a true
minima. Moreover we will study wave functionals such that the
scalar field produces a transition between two minima of the
potential.

One can use these wave functionals in order to compute the
transition probability with the standard interpretation that these
transitions are due to a tunneling between the two minima of the
potential involved in the transition. In order to be more precise,
in the semi-classical approximation the probability to produce a
transition between two vacua at $\phi_{A}$ and $\phi_{B}$ is
the decay rate which can be written as

\begin{multline}\label{TransProOr}
P(A\to B)=\left|\frac{\Psi(\varphi^I_{0},\phi_{B};\varphi^I_{m},\phi_{A})}{\Psi(\varphi^I_{0},\phi_{A};\varphi^I_{m},\phi_{A})}\right|^2
    \\ =\left|\frac{\beta
        e^{\frac{i}{\hbar}S_{0}(\varphi^I_{0},\phi_{B};\varphi^I_{m},\phi_{A})}+\chi
        e^{-\frac{i}{\hbar}S_{0}(\varphi^I_{0},\phi_{B};\varphi^I_{m},\phi_{A})}}{\beta
        e^{\frac{i}{\hbar}S_{0}(\varphi^I_{0},\phi_{A};\varphi^I_{m},\phi_{A})}+\chi
        e^{-\frac{i}{\hbar}S_{0}(\varphi^I_{0},\phi_{A};\varphi^I_{m},\phi_{A})}}\right|^2=\left|e^{-\Gamma}\right|^2,
\end{multline}
where $\varphi^I$ denotes all other fields defined on the superspace
except the scalar field,
$\Psi(\varphi^I_{0},\phi_{B},\varphi^I_{m},\phi_{A})$ is the wave
functional associated to the path which starts in 
$\varphi^I(s=0)=\varphi^I_{0}$, and where the scalar field takes the
value $\phi_{B}$. Moreover the path ends in
$\varphi^I(s=s_{M})=\varphi^I_{m}$, where the scalar field is
denoted by $\phi_{A}$. Furthermore $\beta$ and $\chi$ are the constants
of the linear superposition. In the previous equation we will
consider just the dominant contribution of the exponential terms. Then in
the WKB approximation at first order $\Gamma$ yields
\begin{equation}\label{DefGamma}
    \pm\Gamma=\frac{i}{\hbar}S_{0}(\varphi^I_{0},\phi_{B};\varphi^I_{m},\phi_{A})-\frac{i}{\hbar}S_{0}(\varphi^I_{0},\phi_{A};\varphi^I_{m},\phi_{A}) ,
\end{equation}
where the choice of the signs $\pm$ indicates the dominant terms in
the expression (\ref{TransProOr}). Thus we finally arrive at the
transition probability given by
\begin{multline}\label{TransProb}
P(A\to B)=\exp\left[-2{\rm Re}(\Gamma)\right]
    \\ = \exp\left\{\pm2{\rm Re}\left[\frac{i}{\hbar}S_{0}(\varphi^I_{0},\phi_{B};\varphi^I_{m},\phi_{A})-\frac{i}{\hbar}S_{0}(\varphi^I_{0},\phi_{A};\varphi^I_{m},\phi_{A})\right]\right\} .
\end{multline}

It is worth mentioning that the formalism developed in
\cite{Cespedes:2020xpn,LVTAU}, originally for GR, is general enough
to include other gravitational theories, since it only depends on a
Hamiltonian constraint written in the general form
(\ref{HamConstraint}). In the next sections we will show that it can
perfectly include higher derivative generalizations of GR as the HL
theory. It would be interesting to study at what extent this
formalism can be used for more general theories.

\section{Wheeler-DeWitt equation for Ho\v{r}ava-Lifshitz gravity coupled to matter}
\label{S-HL}

In this section we will discuss the action in HL theory as well as
the action that consider the coupling to a scalar field. We will
consider a metric describing a FLRW universe and a scalar field
depending on the time variable as well as the spatial variables, and
we will obtain the WDW equation for such a system. Although in the
context of cosmology it is usual to use a time dependent field only,
in this case we will allow the scalar field to depend also on the
spatial variables since the anisotropic scaling of both sets of
variables is a key ingredient for the theory. This type of
dependence has been used previously in the context of cosmology for
HL. For example it was used in Ref. \cite{Mukohyama:2009gg} to study
perturbations coming from a scalar field. However, for completeness
and correspondence with the cosmological models, we will also study
the case when the field will only depend on the time coordinate in
Section \ref{S-TT}.

Let us begin by considering the gravitational part of the general
action in projectable HL gravity without a cosmological constant and
without detailed balance. This action can be written as
\cite{Bertolami:2011ka,Sotiriou:2009gy,Sotiriou:2009bx}
    \begin{multline}\label{HLGraviAction}
    S_{\rm HL}=\frac{M^2_{p}}{2}\int dtd^3x N\sqrt{h}\left[K^{ij}K_{ij}-\lambda K^2+R-\frac{1}{M^2_{p}}\left(g_{2}R^2+g_{3}R_{ij}R^{ij}\right) \right. \\ \left. -\frac{1}{M^4_{p}}\left(g_{4}R^3+g_{5}R\left(R_{ij}R^{ij}\right)+g_{6}R^{i}_{j}R^{j}_{k}R^{k}_{i}+g_{7}RD^2R+g_{8}D_{i}R_{jk}D^{i}R^{jk}\right)\right] ,
    \end{multline}
where $N$ is the lapse function, $R_{ij}$ the Ricci tensor with
$i,j=1,2,3$ the spatial indices, $R$ the Ricci scalar, $K_{ij}$ the
extrinsic curvature, $M_{p}$ the Planck mass, $D$ denotes covariant
derivative with respect to the three-metric $h_{ij}$ and $h$ denotes
its determinant, all $g_{n}$ ($n=2,...,8$) are positive
dimensionless running coupling constants and the parameter $\lambda$
runs under the renormalization group flow. GR is in principle
obtained in the limit $\lambda\to 1$ and $g_{n}\to 0$. However this
is not actually fulfilled because of the perturbative IR instability
and the presence of an unphysical degree of freedom as mentioned in the
introduction.

Let us take the FLRW metric with positive curvature that describes a
closed homogeneous and isotropic universe. This metric is written as
    \begin{equation}\label{FLRWMetric}
        ds^2=-N^2(t)dt^2+a^{2}(t)\left[dr^2+\sin^2r\left(d\theta^2+\sin^2\theta d\psi^2\right)\right] ,
    \end{equation}
where as usual $0\leq r\leq\pi$, $0\leq\theta\leq\pi$ and $0\leq\psi\leq 2\pi$. In the context of the ADM formalism, we note that for this metric
    \begin{equation}\label{FLRWADM}
        N=N(t) , \hspace{0.5cm} N_{i}=0 , \hspace{0.5cm} (h_{ij})=a^2(t){\rm diag}\left(1,\sin^2r,\sin^2r\sin^2\theta \right) ,
    \end{equation}
and therefore we will work with a projectable version of HL gravity.
Substituting these values in the action (\ref{HLGraviAction}) we
obtain that for this metric the gravitational action reads
    \begin{multline}\label{FLRWActionGrav}
    S_{\rm HL}=2\pi^2\int dt N\left[-\frac{3(3\lambda-1)M^2_{p}a}{2N^2}\dot{a}^2+3M^2_{p}a-\frac{6}{a}(3g_{2}+g_{3})\right. \\ \left. -\frac{12}{a^3M^2_{p}}(9g_{4}+3g_{5}+g_{6})\right] ,
    \end{multline}
where a dot stands for the derivative with respect to the time
variable. The integral for the spatial slice has been performed,
that is
    \begin{equation}
        {\rm Vol}(X)=\int_{r=0}^{\pi}\int_{\theta=0}^{\pi}\int_{\psi=0}^{2\pi}\sin^2r\sin\theta drd\theta d\psi = 2\pi^2.
    \end{equation}

In order to couple a scalar field $\phi(t,x^{i})$ (where $x^{i}$
denotes collectively the three spatial variables) to this theory we
need to consider actions that are compatible with the anisotropic
scaling symmetries of the theory and UV renormalizability. In fact,
the general scalar action in HL gravity  is found to contain up to 6
order derivatives. This action is written in the form
\cite{Kiritsis:2009sh}
    \begin{equation}\label{ActionMatter}
    S_{m}=\frac{1}{2}\int dtd^3x \sqrt{h}N\left[\frac{(3\lambda-1)}{2N^2}\left(\dot{\phi}-N^i\partial_{i}\phi\right)^2+F(\phi)\right] ,
    \end{equation}
where the function $F(\phi)$ is given by
    \begin{equation}\label{DefinitionF}
    F(\phi)=\phi\left(c_{1}\Delta\phi-c_{2}\Delta^2\phi+c_{3}\Delta^3\phi\right)-V(\phi) ,
    \end{equation}
with $\Delta$ denoting the three-metric laplacian and $V(\phi)$ is
the potential for the scalar field. The constant $c_{1}$ is the
velocity of light in the IR limit, whereas the two other constants
are related to the energy scale $M$ as
    \begin{equation}
        c_{2}=\frac{1}{M^2} , \hspace{0.5cm} c_{3}=\frac{1}{M^4} .
    \end{equation}
There are three more possible terms that can be part of
(\ref{DefinitionF}) constructed as products of derivatives, but we
restrict ourselves to the terms just described.

Since the three-metric derived from the FLRW metric is just a scale
factor times the metric of the three-sphere ${\bf S}^{3}$, we can
use the spherical harmonic functions defined in this space to expand
our scalar functions \cite{Lindblom:2017maa,Sandberg:1978}. These
functions can be defined in our spatial three-metric as
eigenfunctions of the laplacian of the form
    \begin{equation}\label{SpHarDeg}
    \Delta Y_{nlm}(x^{i})=-\frac{n(n+2)}{a^2}Y_{nlm}(x^{i}),
    \end{equation}
where $n$ is an integer. They obey the orthonormality condition
    \begin{equation}\label{OrtoCondiO}
    \frac{1}{a^3}\int \sqrt{h}Y_{nlm}(x^{i})Y^{*}_{n'l'm'}(x^{i})d^3x=\delta_{nn'}\delta_{ll'}\delta_{mm'} .
    \end{equation}
Since these functions form a complete basis, we can expand any
scalar function defined on the sphere in terms of them as
    \begin{equation}\label{ExpansionScalarFunc}
     f(x^{i})=\sum_{n=0}^{\infty}\sum_{l=0}^{n}\sum_{m=-l}^{l}\alpha_{nlm}Y_{nlm}(x^{i})=\sum_{\{n,l,m\}}\alpha_{nlm}Y_{nlm}(x^{i}) .
    \end{equation}
Therefore, using this basis, we can expand the scalar field as
    \begin{equation}\label{ExpansionScalarField}
    \phi(t,x^{i})=\sum_{\{n,l,m\}}\phi_{nlm}(t)Y_{nlm}(x^{i}) ,
    \end{equation}
where the fields $\phi_{nlm}(t)$ are real functions depending only
on the time variable. We can also expand the scalar field potential
as
    \begin{equation}\label{ExpansiconScalarPotential}
        V(\phi)=\sum_{\{n,l,m\}}V_{nlm}(t)Y_{nlm}(x^{i}) ,
    \end{equation}
where the funcions $V_{nlm}(t)$ depends on all the functions
$\phi_{nlm}(t)$ in general. Substituting
(\ref{ExpansionScalarField}) and (\ref{ExpansiconScalarPotential})
back into the action (\ref{ActionMatter}) we obtain that for the
FLRW metric the field part of the action is written as
    \begin{multline}\label{ActionMatterF}
    S_{m}=\sum_{\{n,l,m\}}\frac{1}{2}\int dt \left\{ \frac{3\lambda-1}{2N}a^3\dot{\phi}_{nlm}^2-Na\left[c_{1}\beta_{n}+\frac{c_{2}\beta^2_{n}}{a^2}+\frac{c_{3}\beta^3_{n}}{a^4}\right]\phi_{nlm}^2\right. \\ \left. -Na^3\gamma_{nlm}V_{nlm} \right\} ,
    \end{multline}
where $\beta_{n}=n(n+2)$ and
        \begin{equation}\label{DefGammasAux}
        \gamma_{nlm}=\int_{r=0}^{\pi}\int_{\psi=0}^{2\pi}\int_{\theta=0}^{\pi}\sin^2r\sin\theta Y_{nlm}(r,\theta,\psi)drd\psi d\theta ,
        \end{equation}
are constants. Finally, by considering together both actions
(\ref{FLRWActionGrav}) and (\ref{ActionMatterF}) we obtain that the
full lagrangian describing HL gravity coupled to a scalar field is
    \begin{multline}\label{LagrangianFull}
    \mathcal{L}=2\pi^2N\left[-\frac{3M^2_{p}\dot{a}^2a}{2N^2}(3\lambda-1)+3M^2_{p}a-\frac{6}{a}(3g_{2}+g_{3})-\frac{12}{a^3M^2_{p}}(9g_{4}+3g_{5}+g_{6})\right] \\ +\sum_{\{n,l,m\}}\left\{ \frac{3\lambda-1}{4N}a^3\dot{\phi}_{nlm}^2-\frac{Na}{2}\left[c_{1}\beta_{n}+\frac{c_{2}\beta^2_{n}}{a^2}+\frac{c_{3}\beta^3_{n}}{a^4}\right]\phi_{nlm}^2-\frac{Na^3}{2}\gamma_{nlm}V_{nlm} \right\} .
    \end{multline}
We have in this case that the degrees of freedom are the fields
$\left\{a,\phi_{nlm}\right\}$. Their canonical momenta turn out to
be
    \begin{equation}\label{CanonicalMomenta}
    \pi_{a}=-\frac{6\pi^2M^2_{p}(3\lambda-1)}{N}a\dot{a} , \hspace{1cm} \pi_{\phi_{nlm}}=\frac{3\lambda-1}{2N}a^3\dot{\phi}_{nlm} ,
    \end{equation}
and, as it is usual, the lapse function is non-dynamical since
$\pi_{N}=0$. Therefore, we obtain that the Hamiltonian constraint
takes the form
    \begin{multline}\label{HamiltonianConstraint}
    H=N\left\{-\frac{\pi^2_{a}}{12\pi^2M^2_{p}(3\lambda-1)a}+\sum_{\{n,l,m\}}\frac{\pi^2_{\phi_{nlm}}}{(3\lambda-1)a^3}\right. \\ \left. +2\pi^2\left[-3M^2_{p}a+\frac{6}{a}(3g_{2}+g_{3})+\frac{12}{M^2_{p}a^3}(9g_{4}+3g_{5}+g_{6})\right]\right.\\\left.+\frac{1}{2}\sum_{\{n,l,m\}}\left[\left(c_{1}\beta_{n}+\frac{c_{2}}{a^2}\beta^2_{n}+\frac{c_{3}}{a^4}\beta^3_{n}\right)a\phi^2_{nlm}+a^3\gamma_{nlm}V_{nlm}\right]\right\} \simeq 0 .
    \end{multline}

\section{Vacuum transitions in Ho\v{r}ava-Lifshitz gravity}
\label{S-Trans}

Now that we have obtained the Hamiltonian constraint of the HL
gravity coupled to a scalar field depending on all spacetime
variables, let us study the probability transition between two vacua
of the scalar field potential. We note that the form of the
Hamiltonian constraint (\ref{HamiltonianConstraint}) as obtained in
the previous section is of the same general form as the one
considered in (\ref{HamConstraint}) taking the coordinates on
superspace to be $\left\{a,\phi_{nlm}\right\}$. The inverse metric
is given by
    \begin{equation}
    (G^{MN})={\rm diag}\left(-\frac{1}{6\pi^2M^2_{p}(3\lambda-1)a},\frac{2}{(3\lambda-1)a^3}\mathbf{1}_{nlm}\right) ,
    \end{equation}
where $\mathbf{1}_{nlm}$ denotes a vector with length equal to all
the possible values that the set $\{n,l,m\}$ can have and with $1$
in all its entries. In this case then we also have
    \begin{multline}
    f(a,\phi_{nlm},V_{nlm})=2\pi^2\left[-3M^2_{p}a+\frac{6}{a}(3g_{2}+g_{3})+\frac{12}{M^2_{p}a^3}(9g_{4}+3g_{5}+g_{6})\right]\\+\frac{1}{2}\sum_{\{n,l,m\}}\left[\left(c_{1}\beta_{n}+\frac{c_{2}}{a^2}\beta^2_{n}+\frac{c_{3}}{a^4}\beta^3_{n}\right)a\phi^2_{nlm}+a^3\gamma_{nlm}V_{nlm}\right] .
    \end{multline}
Therefore, the general procedure to obtain a solution of the WDW
equation presented in Section \ref{S-WDW} is applicable to the WDW
equation obtained after quantizing the Hamiltonian constraint
(\ref{HamiltonianConstraint}) in HL gravity. In order to study
transitions between two vacua of a scalar field potential, we
consider that all fields $V_{nlm}$ appearing in the expansion of the
potential (\ref{ExpansiconScalarPotential}) have the same minima,
namely one false minimum at $\phi^{A}_{nlm}$ and one true minimum at
$\phi^{B}_{nlm}$, and therefore, the two minima of the scalar field
$\phi(t,x^{i})$ comes only from its time dependence. Therefore, the
transition probability in the semi-classical approach between these
two minima is given by Eq. (\ref{TransProb}).

Following \cite{LVTAU} we can choose the parameter $s$ such that for
the interval $[0,\bar{s}-\delta s]$, where $s=0$ is the initial
value, the field remains close to its value at the true minimum
$\phi_{B}$, and for the interval $[\bar{s}+\delta s,s_{m}]$ the
field remains very close to its value at the false minimum
$\phi_{A}$, that is, we choose the parameter $s$ such that
        \begin{equation}\label{ChooseS}
        \phi(s) \approx
        \begin{cases}
        \phi_{B} , & 0<s<\bar{s}-\delta s,\\
        \phi_{A} , & \bar{s}+\delta s<s<s_{M}.
        \end{cases}
        \end{equation}
But in this case, taking the expansion (\ref{ExpansionScalarField})
and since the spherical harmonics are an orthonormal set, the latter
implies that
    \begin{equation}\label{ChooseSField}
    \phi_{nlm}(s) \approx
    \begin{cases}
    \phi_{nlm}^{B} , & 0<s<\bar{s}-\delta s,\\
    \phi_{nlm}^{A} , & \bar{s}+\delta s<s<s_{M} ,
    \end{cases}
    \end{equation}
and similarly for the potentials
    \begin{equation}\label{ChooseSPotential}
    V_{nlm}(s) \approx
    \begin{cases}
    V_{nlm}^{B} , & 0<s<\bar{s}-\delta s,\\
    V_{nlm}^{A} , & \bar{s}+\delta s<s<s_{M} .
    \end{cases}
    \end{equation}
Therefore, using the general form of the action
(\ref{ClassActionGen}) we obtain in this case
    \begin{multline}\label{ClassicalAction1}
    S_{0}\left(a_{0},\phi^{B}_{nlm};a_{m},\phi^{A}_{nlm}\right)=-4\pi^2\left[\int_{0}^{\bar{s}-\delta s}\frac{ds}{C(s)}f\big\rvert_{\phi_{nlm}=\phi^{B}_{nlm}}+\int_{\bar{s}-\delta s}^{\bar{s}+\delta s}\frac{ds}{C(s)}f\right. \\ \left. +\int_{\bar{s}+\delta s}^{s_{m}}\frac{ds}{C(s)}f\big\rvert_{\phi_{nlm}=\phi^{A}_{nlm}}\right] ,
    \end{multline}
and
    \begin{equation}\label{ClassicalAction2}
    S_{0}\left(a_{0},\phi^{A}_{nlm};a_{m},\phi^{A}_{nlm}\right)=-4\pi^2\int_{0}^{s_{m}}\frac{ds}{C(s)}f\big\rvert_{\phi_{nlm}=\phi^{A}_{nlm}}
    .
    \end{equation}
Consequently, the logarithm of the probability (\ref{DefGamma}) is
given in this case by
    \begin{multline}\label{GammaF1}
    \pm\Gamma=\frac{i}{\hbar}\left[-4\pi^2\int_{0}^{\bar{s}-\delta s}\frac{ds}{C(s)}f\big\rvert_{\phi_{nlm}=\phi^{B}_{nlm}}+4\pi^2\int_{0}^{\bar{s}-\delta s}\frac{ds}{C(s)}f\big\rvert_{\phi_{nlm}=\phi^{A}_{nlm}}\right. \\
    \left.-4\pi^2\int_{\bar{s}-\delta s}^{\bar{s}+\delta s}\frac{ds}{C(s)}\left\{\frac{1}{2}\sum_{\{n,l,m\}}\left[a\left(c_{1}\beta_{n}+\frac{c_{2}}{a^2}\beta^2_{n}+\frac{c_{3}}{a^4}\beta^3_{n}\right)\left(\phi^2_{nlm}-(\phi^{A}_{nlm})^2\right)\right.\right.\right. \\
    \left. \left.\left. +a^3\gamma_{nlm}(V_{nlm}-V^{A}_{nlm})\right]\right\}\right] .
    \end{multline}

We note that the last term of Eq. (\ref{GammaF1}) can be written as
    \begin{equation}\label{NewPotentials0}
    -4\pi^2i\int_{\bar{s}-\delta s}^{\bar{s}+\delta s}\frac{ds}{C(s)}\left[\frac{1}{2}\sum_{\{n,l,m\}}a^3\gamma_{n,l,m}(V_{n,l,m}-V^{A}_{nlm})\right]=-4\pi^2i\int_{\bar{s}-\delta s}^{\bar{s}+\delta s}\frac{ds}{C(s)}a^3\left[V_{0}-V_{0}^{A}\right] ,
    \end{equation}
with a potential defined by
    \begin{equation}\label{DefPot0}
    V_{0}=\frac{1}{2}\sum_{\{n,l,m\}}\gamma_{nlm}V_{nlm} .
    \end{equation}
We note that this term has the same form as the one considered in
Refs. \cite{Cespedes:2020xpn,LVTAU} regarding the portion of the
integral in which the scalar field can vary, therefore we can also
interpret this term as a tension term taking
    \begin{equation}\label{DefTension0}
    2\pi^2\bar{a}^3T_{0}=-4\pi^2i\int_{\bar{s}-\delta s}^{\bar{s}+\delta s}\frac{ds}{C(s)}a^3\left[V_{0}-V_{0}^{A}\right] .
    \end{equation}
Moreover, we note that the term that contains $c_{1}$ in
(\ref{GammaF1}) can be written as \small
    \begin{equation}\label{NewPotential1}
    -4\pi^2i\int_{\bar{s}-\delta s}^{\bar{s}+\delta s}\frac{ds}{C(s)}\left[\frac{1}{2}\sum_{\{n,l,m\}}ac_{1}\beta_{n}\left(\phi^2_{nlm}-(\phi^{A}_{nlm})^2\right)\right]=-4\pi^2i\int_{\bar{s}-\delta s}^{\bar{s}+\delta s}\frac{ds}{C(s)}c_{1}a\left[V_{1}-V_{1}^{A}\right] ,
    \end{equation}
\normalsize
with
    \begin{equation}\label{DefPotential1}
    V_{1}=\frac{1}{2}\sum_{\{n,l,m\}}\beta_{n}\phi_{nlm}^2 .
    \end{equation}
Although this function has no minima in the points considered, it is
a function of the scalar fields that can be interpreted as a new
effective potential with the form of a mass term. Therefore,
applying the same logic used in \cite{LVTAU} to such terms, we can
define a new contribution for the tension term as
    \begin{equation}\label{DefTension1}
    2\pi^2c_{1}\bar{a}T_{1}=-4\pi^2i\int_{\bar{s}-\delta s}^{\bar{s}+\delta s}\frac{ds}{C(s)}c_{1}a\left[V_{1}-V_{1}^{A}\right] .
    \end{equation}
Similarly, we define for the two remaining terms
    \begin{equation}\label{DefPotentials23}
    V_{2}=\frac{1}{2}\sum_{\{n,l,m\}}\beta_{n}^2\phi_{nlm}^2 , \hspace{1cm}
    V_{3}=\frac{1}{2}\sum_{\{n,l,m\}}\beta_{n}^3\phi_{nlm}^2 ,
    \end{equation}
and the two contributions to the tension terms as
    \begin{equation}\label{DefTension2}
    2\pi^2\frac{c_{2}}{\bar{a}}T_{2}=-4\pi^2i\int_{\bar{s}-\delta s}^{\bar{s}+\delta s}\frac{ds}{C(s)}\frac{c_{2}}{a}\left[V_{2}-V_{2}^{A}\right]
    ,
    \end{equation}
    \begin{equation}\label{DefTension3}
    2\pi^2\frac{c_{3}}{\bar{a}^{3}}T_{3}=-4\pi^2i\int_{\bar{s}-\delta s}^{\bar{s}+\delta s}\frac{ds}{C(s)}\frac{c_{3}}{a^3}\left[V_{3}-V_{3}^{A}\right] .
    \end{equation}

On the other hand, in order to do the two first integrals in
(\ref{GammaF1}) where all scalar fields are constants we use the
general solutions (\ref{CsGeneral}) and (\ref{DerivSolGen}), then
after changing the integration variables from $s$ to $a$ according
to $ds=\left(\frac{da}{ds}\right)^{-1}da$ we obtain
    \begin{multline}\label{IntegralConst}
    -4\pi^2\int_{0}^{\bar{s}-\delta s}\frac{ds}{C(s)}f\big\rvert_{\phi_{nlm}=\phi^{A,B}_{nlm}} \\ =\pm 4\pi^3M_{p}\sqrt{3(3\lambda-1)}\int_{a_{0}}^{\bar{a}-\delta a}\sqrt{-\alpha^{A,B}_{1}a^2+\alpha^{A,B}_{2}+\frac{\alpha^{A,B}_{3}}{a^2}+V^{A,B}_{0}a^4}da ,
    \end{multline}
where
    \begin{equation}\label{DefConstIntegralO}
    \begin{split}
    \alpha^{A.B}_{1}&=6\pi^2M^2_{p}-c_{1}V^{A,B}_{1}  ,\\ \alpha^{A,B}_{2}&=12\pi^2(3g_{2}+g_{3})+c_{2}V^{A,B}_{2} , \\ \alpha^{A,B}_{3}&=\frac{24\pi^2}{M^2_{p}}(9g_{4}+3g_{5}+g_{6})+c_{3}V^{A,B}_{3} .
    \end{split}
    \end{equation}
Therefore, substituting Eqs. (\ref{DefTension0}),
(\ref{DefTension1}), (\ref{DefTension2}), (\ref{DefTension3}) and
(\ref{IntegralConst}) back into (\ref{GammaF1}) we obtain
    \begin{multline}\label{GammaF2}
    \pm\Gamma=\pm\frac{ 4\pi^3M_{p}\sqrt{3(3\lambda-1)}}{\hbar}\left[\int_{a_{0}}^{\bar{a}-\delta a}F\left(\alpha^{B}_{1},\alpha^{B}_{2},\alpha^{B}_{3},V^{B}_{0},a\right)da \right. \\ \left. -\int_{a_{0}}^{\bar{a}-\delta a}F\left(\alpha^{A}_{1},\alpha^{A}_{2},\alpha^{A}_{3},V^{A}_{0},a\right)da \right]+\frac{2\pi^2}{\hbar}\left[\bar{a}^3T_{0}+c_{1}\bar{a}T_{1}+\frac{c_{2}}{\bar{a}}T_{2}+\frac{c_{3}}{\bar{a}^3}T_{3}\right] .
    \end{multline}
where we have defined the function
    \begin{equation}\label{FDef}
        F(a,b,c,e,x)=\sqrt{a x^2-b-\frac{c}{x^2}-e x^4} ,
    \end{equation}
and it is worth noting that in the above result the sign ambiguity
on the left hand side comes from the arguments leading to Eq.
(\ref{DefGamma}) whereas the one on the right comes from the fact
that the general solution (\ref{CsGeneral}) and (\ref{DerivSolGen})
for the system of equations gives a solution for $C^2(s)$ wich
produces a sign ambiguity in Eq. (\ref{IntegralConst}). Therefore
both ambiguities are independent to each other.

As it is well known the IR limit of HL gravity for an FLRW metric is
achieved in the limit $\lambda\to1$ and $a>>1$, and corresponds to
GR with an extra degree of freedom  albeit with the instability
problems mentioned in the introduction section. We note that the
kinetic term for the scale factor in (\ref{LagrangianFull}) is
    \begin{equation}
    -2\pi^2\left[\frac{3M^2_{p}a\dot{a}^2}{2N}(3\lambda-1)\right] .
    \end{equation}
In the GR case considered in \cite{LVTAU} it is given by
    \begin{equation}
    -\frac{3a\dot{a}^2}{N} ,
    \end{equation}
because the $2\pi^2$ term in that case is a global multiplicative
factor to the full lagrangian and therefore it can be ignored. Thus,
in order to obtain the same kinetic term in both cases in the limit
$\lambda\to1$, we consider units such that $2\pi^2M^2_{p}=1$. This
choice of units will allow us to compare directly the transition
probability to the one obtained in GR.

Then, we finally obtain for the logarithm of the transition probability
    \begin{multline}\label{GammaFF}
    \pm\Gamma=\pm\frac{ 2\pi^2\sqrt{6(3\lambda-1)}}{\hbar}\left[\int_{a_{0}}^{\bar{a}-\delta a}F\left(\alpha^{B}_{1},\alpha^{B}_{2},\alpha^{B}_{3},V^{B}_{0},a\right)da \right. \\ \left. -\int_{a_{0}}^{\bar{a}-\delta a}F\left(\alpha^{A}_{1},\alpha^{A}_{2},\alpha^{A}_{3},V^{A}_{0},a\right)da
    \right]+\frac{2\pi^2}{\hbar}\left[\bar{a}^3T_{0}+c_{1}\bar{a}T_{1}+\frac{c_{2}}{\bar{a}}T_{2}+\frac{c_{3}}{\bar{a}^3}T_{3}\right]
    \end{multline}
with
    \begin{equation}\label{DefConstIntegral}
    \begin{split}
    \alpha^{A.B}_{1}&=3-c_{1}V^{A,B}_{1}  ,\\ \alpha^{A,B}_{2}&=12\pi^2(3g_{2}+g_{3})+c_{2}V^{A,B}_{2} , \\ \alpha^{A,B}_{3}&=48\pi^3(9g_{4}+3g_{5}+g_{6})+c_{3}V^{A,B}_{3} .
    \end{split}
    \end{equation}
We note that in contrast to the results obtained using GR for all
the types of metrics considered in \cite{LVTAU} this transition
probability is described by five parameters, and by extremizing the
latter with respect to $\bar{a}$ we can at most reduce them by one.
It is also important to note that the above integrals cannot be done
explicitly for any values of the $\alpha_{i}$ constants.
Nonetheless, it is an expression valid for any value of the
potentials and interestingly it is a general expression that does
not depend on having to consider the different modes contributing to
the expansion in Eq. (\ref{ExpansionScalarField}) separately.

As it is explained in \cite{Cespedes:2020xpn,LVTAU} the choice of
$s$ as in (\ref{ChooseS}) is useful to obtain exact solutions for
the transition probabilities that leads to the same solutions as the
ones obtained using euclidean methods. But there is also room to
consider $s$ in different ways. For example, we can also choose $s$
as the distance in field space. This choice allows us to show that
we can have classical transitions just because the metric in
superspace for the WDW equation considered here coming from the
Hamiltonian constraint (\ref{HamiltonianConstraint}) is non-positive
definite, as is the case for all the metrics considered in
\cite{LVTAU}.

Now that we have computed the transition probability for HL gravity
in general, let us consider its two limits of importance, namely,
the infrared and the ultraviolet limit. The first enable us to
compare directly with the result found by using GR and the latter
allows us to highlight the contributions for high energies that
marks the importance of HL gravity.

Taking the IR limit of (\ref{GammaFF}), that is, taking
$\lambda\to1$ and $a>>1$, we obtain
    \begin{multline}\label{GammaIR}
    \pm\Gamma_{IR}=\mp \frac{4\pi^2}{\hbar}\sqrt{\frac{1}{3}}\left[\frac{(\alpha^{B}_{1})^{3/2}}{V^{B}_{0}}\left(1-\frac{V^{B}_{0}}{\alpha^{B}_{1}}a^2\right)^{3/2}\bigg\rvert_{a_{0}}^{\bar{a}-\delta a}-\frac{(\alpha^{A}_{1})^{3/2}}{V^{A}_{0}}\left(1-\frac{V^{A}_{0}}{\alpha^{A}_{1}}a^2\right)^{3/2}\bigg\rvert_{a_{0}}^{\bar{a}-\delta a}\right]\\ +\frac{2\pi^2}{\hbar}\left[\bar{a}^3T_{0}+c_{1}\bar{a}T_{1}\right] .
    \end{multline}
Therefore, we find in the infrared an expression quite similar to
the GR result plus one degree of freedom extra coming from the
$c_{1}$ term in the action for the scalar field
(\ref{ActionMatterF}) as expected. In order to compare directly our
result to the result obtain for GR we can for the moment set
$c_{1}=0$, then the last result simplifies to
    \begin{equation}\label{GannaIRF}
    \pm\Gamma_{IR}=\mp \frac{12\pi^2}{\hbar}\left[\frac{1}{V^{B}_{0}}\left(1-\frac{V^{B}_{0}}{3}a^2\right)^{3/2}\bigg\rvert_{a_{0}}^{\bar{a}-\delta a}-\frac{1}{V^{A}_{0}}\left(1-\frac{V^{A}_{0}}{3}a^2\right)^{3/2}\bigg\rvert_{a_{0}}^{\bar{a}-\delta a}\right] +\frac{2\pi^2}{\hbar}\bar{a}^3T_{0},
    \end{equation}
that is the same result obtained for GR in
\cite{Cespedes:2020xpn,LVTAU}. The only difference comes in the
choice of $a_{0}$, since for consistency of the integral
approximation we have here that $a_{0}>>1$, therefore it cannot be
chosen to be zero. Thus, the difference between this result and the
GR one are only constants. We also note that the potential appearing
in this expression is not the potential found originally in the
scalar field action, rather it is an effective potential appearing
after integration of the harmonic functions. Considering the thin wall limit $\delta a\to0$ and extremizing the above result with respect to $\bar{a}$ we obtain
     \begin{equation}\label{FLRWTensionA}
     T_{0}=\pm 2\left(\sqrt{\frac{1}{\bar{a}^2}-\frac{V^{A}_{0}}{3}}-\sqrt{\frac{1}{\bar{a}^2}-\frac{V^{B}_{0}}{3}}\right) .
     \end{equation}
Then substituting it back into Eq. (\ref{GannaIRF}) and choosing the
plus sign in the right-hand side we obtain
    \begin{multline}\label{GammaIRCGRF}
    \pm\Gamma_{IR}= \frac{12\pi^2}{\hbar}\left\{\frac{1}{V^{B}_{0}}\left[\left(1-\frac{V^{B}_{0}}{3}\bar{a}^2\right)^{3/2}-\left(1-\frac{V^{B}_{0}}{3}a_{0}^2\right)^{3/2}\right]\right. \\ \left. -\frac{1}{V^{A}_{0}}\left[\left(1-\frac{V^{A}_{0}}{3}\bar{a}^2\right)^{3/2}-\left(1-\frac{V^{A}_{0}}{3}a_{0}^2\right)^{3/2}\right]-\frac{\bar{a}^2}{3}\left(\sqrt{1-\frac{V^{A}_{0}}{3}\bar{a}^2}-\sqrt{1-\frac{V^{B}_{0}}{3}\bar{a}^2}\right)\right\} .
    \end{multline}
Therefore, in this limit the transition probability is finally
described in terms of just one parameter (considering $a_{0}$ as a
constant).

If we take $a_{0}=0$ in (\ref{GannaIRF}), consider the thin wall
limit and rename $V_{0}\to V$ and $T_{0}\to T$, we obtain  the
result found in \cite{Cespedes:2020xpn,LVTAU} for GR
    \begin{equation}\label{GannaIRFGR}
    \pm\Gamma_{GR}=\mp \frac{12\pi^2}{\hbar}\left[\frac{1}{V_{B}}\left(1-\frac{V_{B}}{3}a^2\right)^{3/2}\bigg\rvert_{a_{0}}^{\bar{a}}-\frac{1}{V_{A}}\left(1-\frac{V_{A}}{3}a^2\right)^{3/2}\bigg\rvert_{a_{0}}^{\bar{a}}\right] +\frac{2\pi^2}{\hbar}\bar{a}^3T ,
    \end{equation}
then, extremizing we obtain
     \begin{equation}\label{FLRWTensionAGR}
    T=\pm 2\left(\sqrt{\frac{1}{\bar{a}^2}-\frac{V_{A}}{3}}-\sqrt{\frac{1}{\bar{a}^2}-\frac{V_{B}}{3}}\right) .
    \end{equation}
Thus, finally the logarithm of the transition probability in GR is
written in terms of just one parameter as\footnote{We note that in \cite{LVTAU} we miswrote the sign of the last term in Eq. (\ref{GammaGR}).}
    \begin{multline}\label{GammaGR}
    \pm\Gamma_{GR}= \frac{12\pi^2}{\hbar}\left\{\frac{1}{V_{B}}\left[\left(1-\frac{V_{B}}{3}\bar{a}^2\right)^{3/2}-1\right]-\frac{1}{V_{A}}\left[\left(1-\frac{V_{A}}{3}\bar{a}^2\right)^{3/2}-1\right]\right. \\ \left. -\frac{\bar{a}^2}{3}\left(\sqrt{1-\frac{V_{A}}{3}\bar{a}^2}-\sqrt{1-\frac{V_{B}}{3}\bar{a}^2}\right)\right\}
    .
    \end{multline}

We can see from (\ref{FLRWTensionAGR}) that since $V_{A}>V_{B}$
choosing the plus sign on the right hand side of (\ref{GannaIRFGR})
implies that $T>0$ always. The same is true regarding $T_{0}$. As we
know from \cite{Cespedes:2020xpn,LVTAU} this choice of sign allows
us to obtain the results found using the euclidean approach in
\cite{Parke:1982pm}. It can be proven that the right hand side of
(\ref{GammaGR}) is always positive and therefore, in order to have a
well defined probability defined by (\ref{TransProb}) we choose the
plus sign in the left hand side as well. We note however from
(\ref{FLRWTensionA}) and (\ref{FLRWTensionAGR}) that in order to
have a well defined tension we need the terms inside the square
roots to be positive. If both potential minima are negative, we see
that this is indeed satisfied for all values of $\bar{a}$. However,
if at least one of the potential minima is positive, we see that the
tension will only be well defined until $\bar{a}$ is big enough,
that is, in this case, the tension term is well defined and
consequently the expressions (\ref{GammaIRCGRF}) and (\ref{GammaGR})
are valid only in an interval from $0$ until an upper bound for
$\bar{a}$.

Let us study now the ultraviolet limit, this is found when $a<<1$.
In this limit Eq. (\ref{GammaFF}) simplifies to
    \begin{equation}\label{GammaUV}
    \pm\Gamma_{UV}=\pm\frac{ 2\pi^2\sqrt{6(3\lambda-1)}}{\hbar}\left[\left(\sqrt{-\alpha^{B}_{3}}-\sqrt{-\alpha^{A}_{3}}\right)\ln a \bigg\rvert_{a_{0}}^{\bar{a}-\delta a}
    \right]+\frac{2\pi^2}{\hbar}\left[\frac{c_{2}}{\bar{a}}T_{2}+\frac{c_{3}}{\bar{a}^3}T_{3}\right].
    \end{equation}
We note however that since all $g_{n}$ are positive, and the $V_{3}$
function defined in (\ref{DefPotentials23}) is also positive for
both minima, the first term is purely imaginary. Therefore, it does
not contribute to the transition probability at all and it can be
ignored. Thus, we finally obtain in this limit
    \begin{equation}\label{GammaUF}
    \pm\Gamma_{UV}=\frac{2\pi^2}{\hbar}\left[\frac{c_{2}}{\bar{a}}T_{2}+\frac{c_{3}}{\bar{a}^3}T_{3}\right] .
    \end{equation}
However this expression does not have an extremal with respect to
$\bar{a}$. Then in this case the probability is described by three
independent parameters. For consistency with the GR result we will
choose the plus sign in the left hand side of the latter equation,
therefore, we note that in order to have a well defined probability
we need the overall sign of the right hand side to be positive.
Thus, we can choose both $T_{2}$ and $T_{3}$ to take always positive
values.

We can see from the GR result (\ref{GannaIRFGR}), or after
extremizing (\ref{GammaGR}), that in any case $P(A\to B)\to1$ when
$\bar{a}\to0$,. However for the UV limit of HL presented in
(\ref{GammaUF}) we have in the contrary $P(A\to B)\to0$ when
$\bar{a}\to0$ and then the probability increases as $\bar{a}$
increases. Therefore in this case the UV behaviour is completely
different for both theories.

Now that we have studied the two limits of interest. We proceed to
compare the full result for the transition probability valid for all
$\bar{a}$ (\ref{GammaFF}) to the GR result. As we have said, for
consistency with GR we are going to choose the plus sign in the left
hand side and the minus sign in the right and use the thin wall
limit, therefore we will consider
    \begin{multline}\label{GammaFFF}
    \Gamma=-\frac{ 2\pi^2\sqrt{6(3\lambda-1)}}{\hbar}\left[\int_{a_{0}}^{\bar{a}}F\left(\alpha^{B}_{1},\alpha^{B}_{2},\alpha^{B}_{3},V^{B}_{0},a\right)da \right. \\ \left. -\int_{a_{0}}^{\bar{a}}F\left(\alpha^{A}_{1},\alpha^{A}_{2},\alpha^{A}_{3},V^{A}_{0},a\right)da \right]+\frac{2\pi^2}{\hbar}\left[\bar{a}^3T_{0}+c_{1}\bar{a}T_{1}+\frac{c_{2}}{\bar{a}}T_{2}+\frac{c_{3}}{\bar{a}^3}T_{3}\right] .
    \end{multline}
If we want to vary (\ref{GammaFFF}) we will obtain an
expression involving the tension terms and the functions
$F(\alpha^{A,B}_{1},\alpha^{A,B}_{2},\alpha^{A,B}_{3},V^{A,B},\bar{a})$.
However, since those functions are defined on terms of square roots
we need the terms inside to be non-negative in order to have well
defined tension terms, that is we need that
    \begin{equation}\label{CondF}
        \alpha^{A,B}_{1}\bar{a}^2-\alpha^{A,B}_{2}-\frac{\alpha^{A,B}_{3}}{\bar{a}^2}-V^{A,B}_{0}\bar{a}^4\geq0 .
    \end{equation}
We note that in the best scenario, the latter expression implies
only a lower bound on $\bar{a}$ coming from the $\alpha_{3}$ term.
In the other cases, it could happen that we obtain a lower and an
upper bound for $\bar{a}$, that is the tension terms would only be
well defined over an specific interval or it could even happen that
the latter expression is not satisfied at all for any value of
$\bar{a}$ and in that case (\ref{GammaFFF}) would not have an
extremal. In any case, we see that the extremizing procedure is very
dependent on the many parameters of the theory and does not allow us
to obtain well defined tension terms in general, in particular we
never have access to the UV region. Therefore, in order to avoid
these difficulties we are going to compare the results obtained
before the extremizing procedure takes place, that is we will use
for the comparison the GR result (\ref{GannaIRFGR}) choosing the
signs already mentioned and the HL result (\ref{GammaFFF}). Since
the HL expression depends on many independent parameters and the
integrals cannot be made for any values of the constants involved,
we are going to evaluate numerically this expression. Since in the
IR limit we saw that after extremizing $T_{0}$ and $T$ are always
positive, we are going to take positive values for these parameters.
On the other hand, in the UV limit we saw that we can take $T_{2}$
and $T_{3}$ also to be positive. Finally in order to obtain a well
defined probability, we are also going to choose positive values for
the remaining free parameter $T_{1}$. Thus, we will take positive
values for all the tension terms.

\begin{figure}[h]
    \centering
    \includegraphics[width=0.9\textwidth]{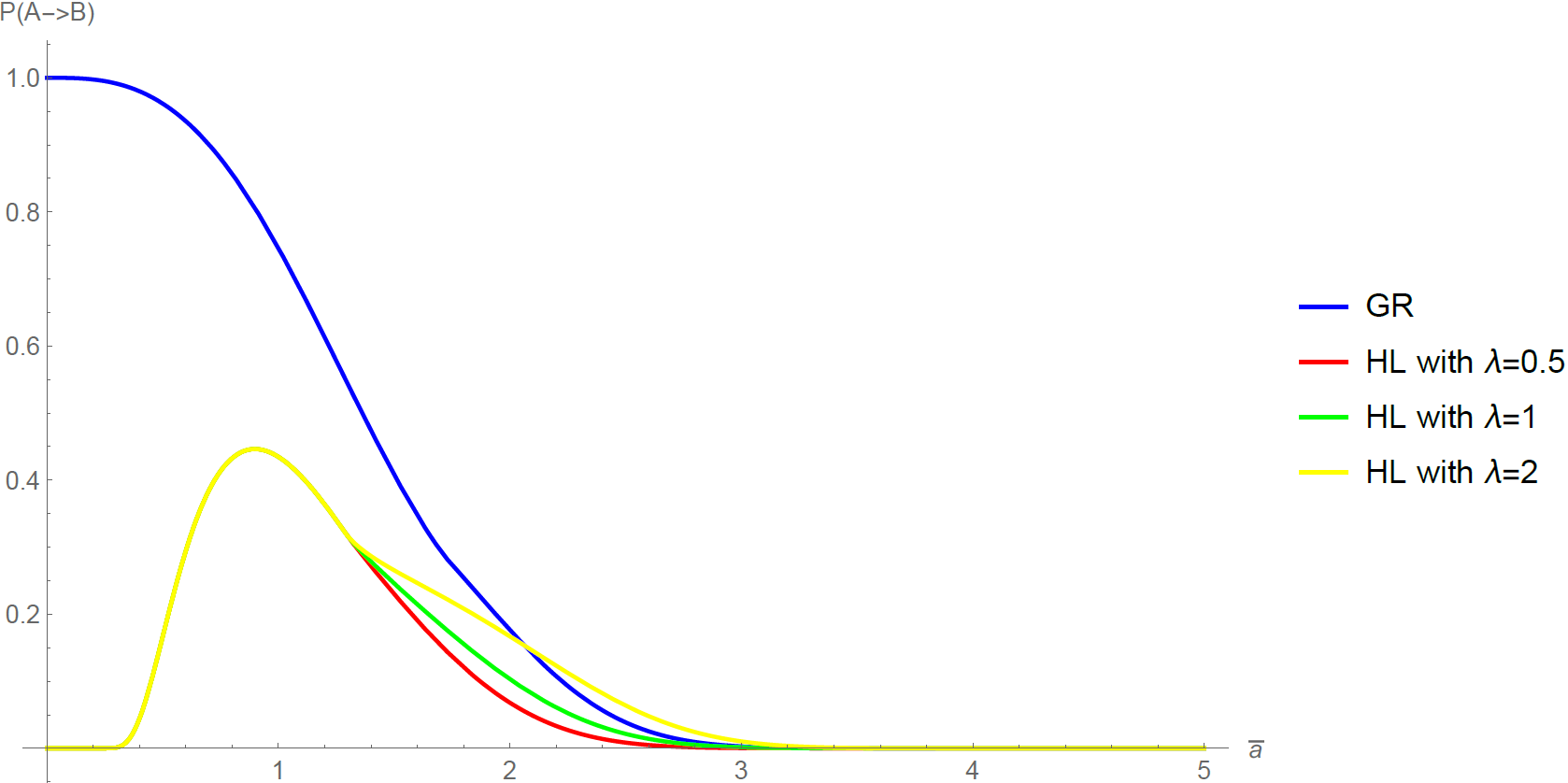}
    \caption{Transition probability in units such that $\frac{24\pi^2}{\hbar}=1$, with $V_{A}=V^{A}_{0}=1$, $V_{B}=V^{B}_{0}=0.1$, $\alpha^{A}_{1}=\alpha^{A}_{2}=\alpha^{A}_{3}=5$,  $\alpha^{B}_{1}=\alpha^{B}_{2}=\alpha^{B}_{3}=4$, $T=T_{0}=2$, $c_{1}T_{1}=c_{2}T_{2}=c_{3}T_{3}=1$, for GR (blue line) and HL with a scalar field depending on all spacetime variables with $\lambda=0.5$ (red line), $\lambda=1$ (green line) and $\lambda=2$ (yellow line). For HL we choose $a_{0}=0.000001$ but the same form is expected for $a_{0}=0$.}
    \label{PlotGeneral}
    \end{figure}

In Figure \ref{PlotGeneral} we show a plot of the transition probabilities
coming from the two theories. We choose units such that
$\frac{24\pi^2}{\hbar}=1$. For the GR result (blue line) we choose
$V_{A}=1$, $V_{B}=0.1$ and $T=2$. We note that in this case the
first term in (\ref{GannaIRFGR}) is negative, therefore, we need to
choose a value for $T$ great enough to obtain a well defined
probability. We see the behaviour outlined earlier, that is, the
probability goes to $1$ in the limit $\bar{a}\to0$ and then it
decreases as $\bar{a}$ increases going to zero. For the HL plots we
choose $V^{A}_{0}=1$, $V^{B}_{0}=0.1$,
$\alpha^{A}_{1}=\alpha^{A}_{2}=\alpha^{A}_{3}=5$,
$\alpha^{B}_{1}=\alpha^{B}_{2}=\alpha^{B}_{3}=4$, $T_{0}=2$ and
$c_{1}T_{1}=c_{2}T_{2}=c_{3}T_{3}=1$, we plot the probability for
three different values of $\lambda$ to see how this parameter
affects the behavior. In this case we choose $a_{0}=0.000001$ in
order to compute the integral numerically, however, we know from the
UV analysis, that $a_{0}=0$ can be chosen without any problem and
the general form will be unaltered. In this case, we also have that
the first term in (\ref{GammaFFF}) is negative and increases with
$\lambda$, therefore, we also need to make sure that the tensions
chosen are big enough to have a well defined probability. This
figure shows the behavior that we described earlier by studying the
different limits of interest. That is, in the IR region the
probability falls in the same manner as the GR result. We note that
$\bar{a}$ has to be big enough so the first term can be positive and
then contribute to the probability, therefore, the different values
of $\lambda$ only affects the curve in the IR region and as
$\lambda$ increases, the probability increases since this term have
the opposite sign that the tension terms. In the UV region the
parameter $\lambda$ has no impact at all and then, everything is
defined by the tension terms. As we have said earlier, the
probability goes to $0$ as $\bar{a}\to0$ and then it increases with
$\bar{a}$. We note that this behavior comes from the  $c_{2}$ and
$c_{3}$ terms, that is, it comes from the extra terms in the action
for the scalar field (\ref{ActionMatter}) and it can be interpreted
as the fact that HL avoids the singularity and predicts these type
of transitions to occur in a UV regime (small $\bar{a}$) but not too
close to the singularity.  Then the probability starts to decrease
and then it goes into the IR behavior just described. We note that
the general form of the plot will be maintained regardless of the
values of the parameters, we only have to make sure that they are
chosen in a way that the tension terms dominates so we can have a
well defined probability. However, specific things as the maximum
height or the point in which both plots match is completely
determined by the parameters and therefore, we cannot say something
about them in general.

In the present section we have obtained a general formula
(\ref{GammaFFF}) for the transition probability  when the scalar
field depends on all spacetime variables. The integrals involved
cannot be done in an analytic form however a numerical computation
was performed and we showed a plot comparing the results with the
one coming from GR. As we pointed out earlier the UV behavior is
found to be completely opposite. This result depends on the extra
terms in the action for the scalar fields, however these terms are
only present if we take a field that depends on the spatial
variables. Therefore, in the next section we are going to study a
scalar field depending only on the time variable.

\textbf{\textit{Phenomenological remarks}}

Before we end this section let us discuss some phenomenological aspects regarding our results. The theory of Ho\v{r}ava-Lifshitz has received a lot of attention and much work has been done since its first proposal from the theoretical as well as the phenomenological point of view. For example in \cite{Dutta:2009jn,Nilsson:2018knn,Nilsson:2021ute} the viability of the different versions of the theory have been tested against various experimental (or observational) data sets coming from different sources such as CMB and BAO collaborations. It is found that the theory is in good agreement with such data and therefore it supports the importance of considering it as a viable theory. It is also interesting to point out that in these works they always work with an FLRW metric with a non-zero curvature since the flat metric gives the same predictions as in General Relativity. Thus the importance of studying such metrics as the one we studied in the present article is also supported by these works.

On the other hand, in the vacuum decay process studied by using the euclidean method, discussed in \cite{ColemanDeLuccia}, the process is described by the nucleation of true vacuum bubbles and its corresponding expansion. This could lead to phenomenological predictions regarding this kind of phase transitions occurring at some point in the evolution of the universe. However, as it was pointed out in \cite{Cespedes:2020xpn} the process studied by using a Hamiltonian approach in the minisuperspace is limited. In fact, in the transition studied there is no notion of bubble nucleation, we can only compare two configurations of three-metrics and then interpret its ratio as a transition probability. Therefore, it is speculated that this formalism is not describing the same process as the euclidean method. It is believed that it may describe a generalization of the tunneling from nothing scenario, that is, we are obtaining probability distributions of creating universes from a tunneling event between two minima of the scalar potential. If we take seriously this interpretation then the scale factor $\bar{a}$ appearing in the expression found for the transition probability would correspond to the value that the scale factor of the created universe would have at the time of creation (its corresponding 'size'). Then, the plot in Figure \ref{PlotGeneral} would tell us that in Ho\v{r}ava-Lifshitz gravity, in the case in which the scalar field depends on the spatial variables, the universe would be created with a scalar factor different from zero and therefore we would avoid the singularity contrary to GR which predicts a singularity at the beginning of the universe. This of course would have potential phenomenological consequences in the physics of the early universe and its corresponding evolution. Therefore, although we are in an speculating phase this kind of transitions are worth studying with more detail.

\section{Transitions for a time dependent scalar field}
\label{S-TT} In the previous sections we studied a scalar field
depending on all coordinates of spacetime and found a transition
probability whose behavior differs completely in the UV regime
comparing to the GR result. However, in cosmology it is more common
to study a scalar field depending only on the time variable as it is
the case in \cite{Kiritsis:2009sh,Tavakoli:2021kyc,Tawfik:2016dvd}.
Therefore, in this section we will consider such a dependence for
the scalar field and study the vacuum transition probability between
two minima of the potential.

In this case, the scalar field action (\ref{ActionMatter}) reduces to
    \begin{equation}\label{AcionMatterTF}
        S_{m}=2\pi^2\int dt a^3(t)\left[\frac{3\lambda-1}{4N^2}\dot{\phi}^2-NV(\phi)\right] ,
    \end{equation}
where we have redefined the scalar field potential appearing in
(\ref{DefinitionF}) as $\frac{V}{2}\to V$ so it coincides with the
usual scalar potential in the action. Since now we have a global
factor of $2\pi^2$ as in the action of the gravitational part
(\ref{FLRWActionGrav}), we can omit this factor. Then, the
lagrangian this time is given by
    \begin{multline}\label{LagrangianFullT}
    \mathcal{L}=N\left[-\frac{3M^2_{p}\dot{a}^2a}{2N^2}(3\lambda-1)+3M^2_{p}a-\frac{6}{a}(3g_{2}+g_{3})-\frac{12}{a^3M^2_{p}}(9g_{4}+3g_{5}+g_{6})\right] \\ +a^3\left[\frac{3\lambda-1}{4N}\dot{\phi}^2-NV\right]  .
    \end{multline}
Therefore we have only two degrees of freedom $a$ and $\phi$, their
canonical momenta are
    \begin{equation}\label{CanonicalMomentaT}
        \pi_{a}=-\frac{3(3\lambda-1)M^2_{p}}{N}a\dot{a} , \hspace{1cm} \pi_{\phi}=\frac{(3\lambda-1)a^3}{2N}\dot{\phi} ,
    \end{equation}
and the Hamiltonian constraint takes the form
    \begin{multline}\label{HamiltonianT}
        H=N\left[\frac{\pi^2_{\phi}}{a^3(3\lambda-1)}-\frac{\pi^2_{a}}{6(3\lambda-1)M^2_{p}a}-3M^2_{p}a+\frac{6}{a}(3g_{2}+g_{3})\right. \\ \left. +\frac{12}{M^2_{p}a^3}(9g_{4}+3g_{5}+g_{6})+a^3V(\phi)\right]\simeq 0 .
    \end{multline}
Comparing this last expression to the general form considered in Eq.
(\ref{HamConstraint}), we note that in this case the coordinates in
superspace are $\{a,\phi\}$ with inverse metric
    \begin{equation}\label{MetricSuperspaceT}
        G^{\phi\phi}=\frac{2}{(3\lambda-1)a^3} , \hspace{0.5cm} G^{aa}=-\frac{1}{3(3\lambda-1)M^2_{p}a} ,
    \end{equation}
and we also have
    \begin{equation}\label{DefinitionfT}
        f(a,\phi)=-3M^2_{p}a+\frac{6}{a}(3g_{2}+g_{3})+\frac{12}{M^2_{p}a^3}(9g_{4}+3g_{5}+g_{6})+a^3V(\phi).
    \end{equation}
In order to study transitions between two minima of the potential,
we choose the parameter $s$ as in expression (\ref{ChooseS}), then
following a similar procedure as in the previous section we obtain
in this case that choosing units such that $M_{p}=1$ (as in the GR
case) and in the thin wall limit  the logarithm of the transition
probability is written as
     \begin{multline}\label{GammaT}
        \pm \Gamma=\pm\frac{2\pi^2\sqrt{6(3\lambda-1)}}{\hbar}\left[\int_{a_{0}}^{\bar{a}}F(3,\bar{\alpha}_{2},\bar{\alpha}_{3},V_{B},a)da-\int_{a_{0}}^{\bar{a}}F(3,\bar{\alpha}_{2},\bar{\alpha}_{3},V_{A},a)da\right]\\ +\frac{2\pi^2}{\hbar}\bar{a}^3T ,
     \end{multline}
where
    \begin{equation}\label{DefAlphasT}
        \bar{\alpha}_{2}=6(3g_{2}+g_{3}) , \hspace{0.5cm} \bar{\alpha}_{3}=12(9g_{4}+3g_{5}+g_{6}) ,
    \end{equation}
the function $F$ is defined in (\ref{FDef}) and as we have mentioned in Section \ref{S-Trans} the sign ambiguities in the last expression are independent.

Now that we have computed the transition probability in general, we
move on to study its behavior in the limiting cases considered
before. For the IR behavior we consider $\lambda\to1$ and $a>>1$ in
the above expression, the result is the same as in (\ref{GannaIRF})
with the same subtlety about $a_{0}$ as discussed in the previous
section. On the other hand in the UV limit we have $a<<1$. However
in this limit we obtain $\Gamma\to0$ as $\bar{a}\to0$. Therefore, we
note that the general behavior of these results is the same as the
GR result in both extreme cases. In fact, we can variate
(\ref{GammaT}) with respect to $\bar{a}$ to obtain
    \begin{equation}\label{TensionT}
        T=\pm\frac{\sqrt{6(3\lambda-1)}}{3\bar{a}^2}\left[F(3,\bar{\alpha}_{2},\bar{\alpha}_{3},V_{A},\bar{a})-F(3,\bar{\alpha}_{2},\bar{\alpha}_{3},V_{B},\bar{a})\right] .
    \end{equation}
Substituting it back in (\ref{GammaT}) we obtain finally
    \begin{multline}\label{GammaTF}
    \pm 2\Re[\Gamma]=\pm\frac{4\pi^2\sqrt{6(3\lambda-1)}}{\hbar}\Re\left[\int_{a_{0}}^{\bar{a}}F(3,\bar{\alpha}_{2},\bar{\alpha}_{3},V_{B},a)da-\int_{a_{0}}^{\bar{a}}F(3,\bar{\alpha}_{2},\bar{\alpha}_{3},V_{A},a)da\right. \\ \left. +\frac{\bar{a}}{3}\left\{F(3,\bar{\alpha}_{2},\bar{\alpha}_{3},V_{A},\bar{a})-F(3,\bar{\alpha}_{2},\bar{\alpha}_{3},V_{B},\bar{a})\right\}\right]
    .
    \end{multline}
Thus, the transition probability is also written in terms of just
one parameter as in the GR result. Therefore, the only difference
between GR and HL in this case is that the transition probability
changes by acquiring two more terms in the square root before
integration  making the integral not possible
to be performed in general and a global factor depending on
$\lambda$ in (\ref{GammaTF}). The qualitative behavior in both the
IR and UV limit is unaltered.

\begin{figure}[h]
        \centering
        \includegraphics[width=0.9\textwidth]{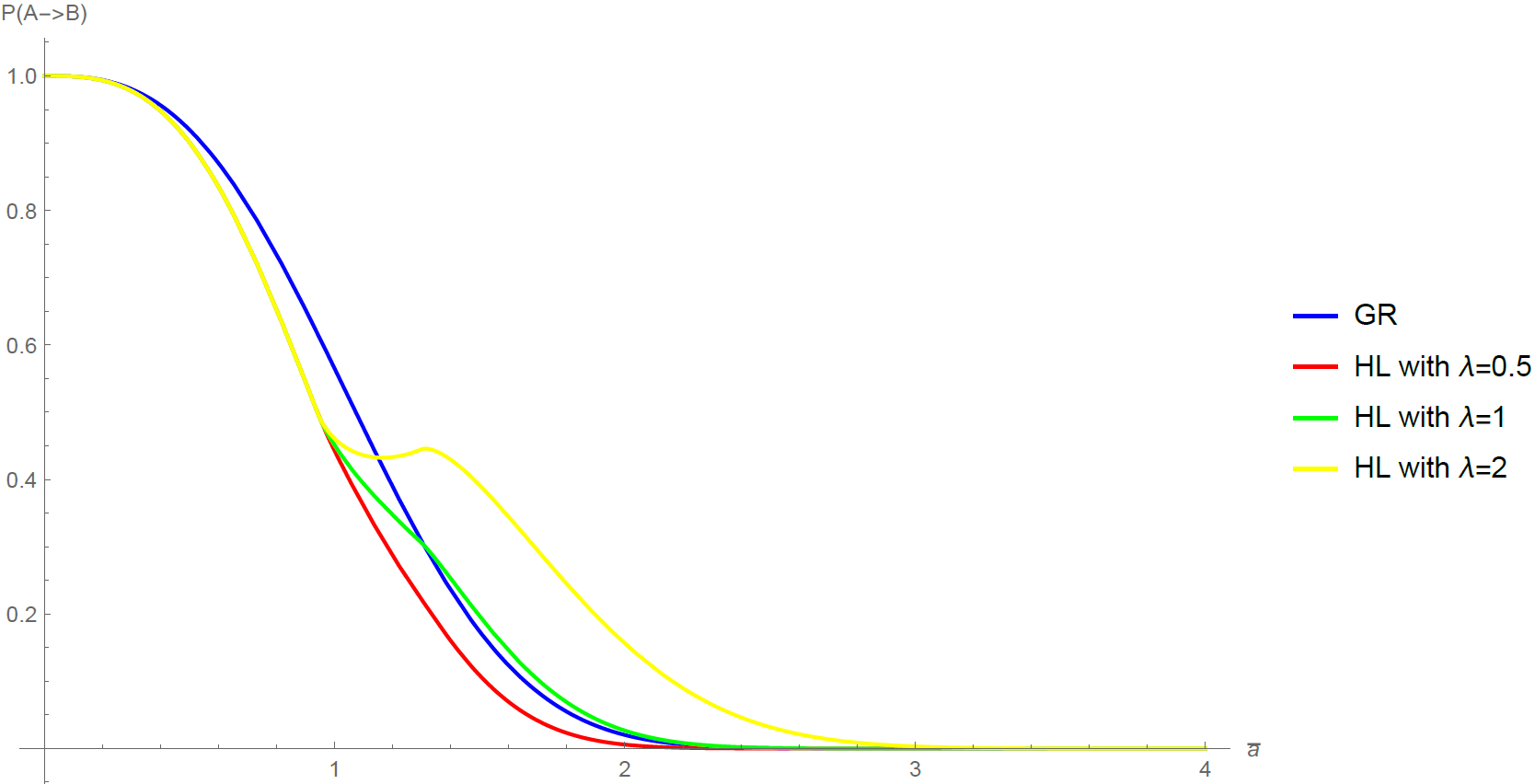}
        \caption{Transition probability in units such that $\frac{24\pi^2}{\hbar}=1$, with $V_{A}=-1$, $V_{B}=-10$, $\bar{\alpha}_{2}=\bar{\alpha}_{3}=5$ and $T=5$, for GR (blue line) and HL with a scalar field depending only on the time variable with $\lambda=0.5$ (red line), $\lambda=1$ (green line) and $\lambda=2$ (yellow line). For HL we choose $a_{0}=0.000001$ but the same form is expected for $a_{0}=0$.}
        \label{PlotTime}
    \end{figure}

In order to compare the result of this section with that of GR in
general, not only on the limiting cases, we note that as in the last
section the extremizing procedure leading to Eq. (\ref{TensionT})
gives rise to some restrictions for the validity of (\ref{GammaTF}).
In particular, it is never well defined when $\bar{a}$ is small.
Therefore, we are going to use the result (\ref{GammaT}) and choose
the minus sign in the right hand side so in the IR limit it
coincides with the GR result and on the left hand side we will
choose the plus sign in accordance with the GR result as well. We
will compare it with the GR result (\ref{GannaIRFGR}) with the sign
choices made in the above section. In both cases we will take the
tension $T$ as an independent positive parameter and take values big
enough so we can have a well defined probability and compute the
integrals numerically. In Figure \ref{PlotTime} we show such
comparison. We choose units such that $\frac{24\pi^2}{\hbar}=1$,
with $V_{A}=-1$, $V_{B}=-10$, $\bar{\alpha}_{2}=\bar{\alpha}_{3}=5$
and $T=5$. For the HL result we show plots for three values of
$\lambda$ and choose $a_{0}=0.000001$ in order to perform a
numerical computation of the integrals, however doing the UV limit
we note that $a_{0}=0$ is possible and it has the same behavior.
This figure shows the limiting behavior that we described earlier,
that is, in the IR and in the UV limits all curves behaves in the
same way, it is in the middle region where their behavior is
modified. In particular we note that in the beginning the HL
probability is smaller than the one of GR  and the contribution of
the $\lambda$ parameter is not noticeable, however when the first
term in (\ref{GammaT}) is big enough, the contribution of the first
term is big enough to separate the curves, and as in the case
considered in the previous section, as $\lambda$ increases the
probability increases. Finally, the three probabilities fall as in
the GR case.

\section{Final Remarks}
\label{S-FinalR} In the present article we have studied the
transition probabilities for an FLRW metric in Ho\v{r}ava-Lifhitz
gravity using a WKB approximation to the WDW equation. The general
procedure proposed in \cite{Cespedes:2020xpn,LVTAU} was found to be
applicable to this case. We used HL theory without detailed balance
and consider an FLRW metric with positive spatial curvature.

We considered two types of scalar fields. First, since the
anisotropic scaling between space and time variables is a key
ingredient of HL theory, we considered a scalar field which depends
on all spacetime variables. This type of dependence is useful to
study cosmological perturbations coming from scalar fields
\cite{Mukohyama:2009gg}. On the other hand, since in cosmology it is
customary to propose an ansatz in which the scalar field depends
only on the time variable, we studied this kind of dependence as
well. For both cases we found analytic expressions for the logarithm
of the transition probabilities in the thin wall limit.

For the scalar field depending on all spacetime variables the
transition probability (\ref{GammaFFF}) was found to depend on five
different parameters coming from the new terms present in the action
for gravity as well as the action from the scalar field in HL
theory. There is only the possibility to reduce just one of these
parameters after an extremizing procedure but such procedure is not
well defined for all values of the scale factor. Taking the IR limit
we found that one degree of freedom extra coming from the scalar
field action survives, which is a common issue regarding the IR
limit of HL theory. However, if we ignore this contribution, we can
obtain a expression that differs from the GR result just by
constants. In the opposite limit, that is, in the UV limit we found
that the probability is described in terms of three independent
parameters and it vanishes in the limit $\bar{a}\to0$. This is
opposite to the GR result in which the probability goes to $1$ in
that limit. We interpret this result as a way in which HL theory
avoids the spatial singularity at $\bar{a}=0$ and predicts these
transitions to occur on the UV regime but away from the singularity.
We note that this behavior comes from the terms in the scalar field
action with spatial derivatives and therefore, it is only possible
in the case in which the scalar field depends on the spatial
variables. In order to visualize these behaviors we plotted the
transition probabilities coming from GR and HL theory. Such plots
were presented in Figure \ref{PlotGeneral}. For the HL results the
integrals involved were done numerically and  we saw that in all
cases the probability begins at zero with $\bar{a}=0$, then it
increases with $\bar{a}$ until at some point it starts to decrease
and then it behaves as in the GR case. We noted that the first
behavior in the UV region is independent of the $\lambda$ parameter
and it is only on the IR where the dependence on this parameter is
noticeable making the probability increase as $\lambda$ increases.

For the scalar field depending only on the time variable the
logarithm of the transition probability found have the same number
of independent parameters as the GR result, that is, after
extremizing we only have one parameter left. However, it also has
dependence on the many constants $g_{n}$ appearing in the extra
terms in the gravity action for HL theory as well as in the
parameter $\lambda$. The behavior of the probability in this case is
found to be the same as the GR result in both the IR as well as the
UV limits. In fact, in the IR limit we obtain the same expression as
the one coming from the scalar field with dependence in all
spacetime variables when we ignore the degree of freedom that
survives this limit and in the UV regime we also have that the
probability goes to $1$ in the limit $\bar{a}\to0$. Therefore, with
a cosmological ansatz, the behavior in the UV regime found by using
GR is unaltered. However in the intermediate region, the probability
is of course modified. In order to visualize the difference in this
region we plotted the transition probabilities coming from GR and HL
theory and showed them in Figure \ref{PlotTime}. In this case we
also carried out a numerical computation of the integrals involved
in the HL result. We noted that at first the probability of HL is
smaller than GR and the contribution from the $\lambda$ parameter is
not noticeable. However, when the scale factor is big enough this
contribution is important and as in the latter case, the probability
increases with $\lambda$. It is interesting to note that using HL
theory instead of GR for a cosmological ansatz of the scalar field
does not have a dramatic change on the transition probability at
least at the semi-classical level we used in this article through
the WKB approximation.

It is worth pointing out that we have used a WKB approximation
and kept only up to first order in the expansion. However, this  level
of semi-classical  approximation is sufficient to obtain the transition probabilities
and we can safely explore the UV regime of both GR as well as HL theory, since the transition 
probabilities are well behaved functions in the UV. It was shown that in the case when the scalar field is only dependent 
on the cosmological time, GR and HL theories give very similar predictions in the WKB approximation. However, the case with a dependence on time and position coordinates for the scalar field, yields very different behavior from the GR case even in the WKB approximation.
It would be interesting to work out higher order contributions from the WKB approximation, which presumably will have the contribution of quantum fluctuations.

It is important to remark as well that we considered closed
universes in the HL theory and obtained well defined transition
probabilities. Therefore, one of the important results obtained
in Ref. \cite{Cespedes:2020xpn} that asserts that this type of
transitions can be carried out keeping the closeness of the spatial
universe can certainly be extended to include the
Ho\v{r}ava-Lifshitz theory of gravity as well.

 \vspace{1cm}
\centerline{\bf Acknowledgments} \vspace{.5cm} D. Mata-Pacheco would
like to thank CONACyT for a grant.



\end{document}